# Single and multiple pin(s)-to-liquid discharges: connecting self-organization patterns and ROS production in liquids for plasma agronomy applications

S. Zhang[1,2], A. Rousseau[1], T. Dufour[1]

[1]*LPP, CNRS, UPMC Univ Paris 06, Ecole polytechnique, Univ. Paris-Sud, Observatoire de Paris, Université Paris-Saclay, Sorbonne Universités, PSL Research University, 4 place Jussieu, 75252 Paris, France*
[2]*Department of Mechanical and Aerospace Engineering, The George Washington University, Washington D.C. 20052 USA*
Corresponding author: shiqiang.zhang@lpp.polytechnique.fr

**Abstract:**
Pin-to-liquid discharges are investigated for the activation of liquids dedicated to agriculture applications. They are characterized through their electrical and optical properties, with a particular attention paid to their filaments and self-organized patterns occurring at the liquid interface. We show how modulating their interaction with ambient air can promote the production of reactive species in liquids such as $H_2O_2$, $NO_2^-$ and $NO_3^-$. The effects of the resulting plasma activated media are reported on lentils seeds.

## 1. Introduction

In plasma medicine as well as plasma agriculture, cold atmospheric discharges are commonly generated following two distinct approaches: (i) a direct approach where substrates or bio-tissues are directly exposed to the plasma and (ii) an indirect approach where cold plasma is utilized to activate a liquid medium subsequently applied on the living system [1-2]. The direct approach allows long/short-lived species and ions transport, UV radiation, gas temperature, transient electric field and gas flow effects while only long-lived species (typically hydrogen peroxide, peroxynitrite, nitrate and nitrite radicals) occur in the indirect approach [3].

The plasma activation of liquids stands for an important issue in plasma agriculture. Although most of the teams working in this field utilize DBD and plasma jets to activate water, we propose here pin-to-liquid discharges as an alternative owing to their simpler design, their lower cost and also the ability to develop multiple pin-to-liquid discharges for the treatment of large liquid areas. In that framework, fundamental questions must be addressed, in particular on the connection between plasma filaments and self-organized patterns on the liquid interface. Previous works have already evidenced patterns formation using DBD, MHCD, jet-like discharges [4,5,6,7] and investigated parameters such as applied voltage, gap distance, exciting frequency, gas composition. DC current and liquid conductivity effects on self-organization have also been investigated using pin-to-liquid glow-like discharges [8]. However, extensive mechanisms on their formation on liquid interfaces are still challenging as well as how their interaction can be modulated with ambient air and hence promote the production of reactive species in the liquid phase. Addressing these questions may be of first interest for applications dealing with the plasma-activation of liquids, in particular in plasma agronomy where concentrated "cocktails" of reactive species are expected at a lower cost.

## 2. Experimental setups

The single pin-to-liquid discharge device is introduced in **Figure 1**. The cathode pin electrode (tungsten, Ø = 2 mm) is supplied by a DC power source (Power Design, model 1570A, 1 – 3012 V, 40 mA) via a ballast resistor (90 kΩ). Below this cathode is placed a glass vessel where the liquid to be treated is contained. This liquid is either mineralized water, demineralized water or non-drinking water. The distance between pin electrode and liquid interface is 2-5 mm. At vessel's bottom is placed the anode plate (thickness=0.2 mm, Ø=22 mm). Besides, a multiple pins-to-liquid discharge has been designed as shown in Figure 1, showing the same previous pins parameters. This matrix configuration is investigated for preparing PAMs with high ROS densities in liquid media.

The electrical properties of the gaseous phase are characterized using voltage probes and Rogowsky coil connected to a LeCroy Wavesurfer 3054 oscilloscope. The discharge optical properties have been determined with an optical emission spectrometer coupled with an intensified CCD camera (iStar 734 from Andor Technologies). Several methods are under study to measure the discharge electric field, CARS laser spectroscopy [9], Stark polarization spectroscopy in presence of helium gas [10], or even Pockels-effect method [11]. Here, the electric field is determined using a ratio of nitrogen bands intensities [12]. The gas temperature is also determined via OES.

The main chemical properties and longlife reactive species of the plasma-activated media are characterized using ion selective electrode probes (Nitrates), chemical probes (pH, redox potential), spectrophotometry (Jenway 7315) and fluorometry (hydrogen peroxide, nitrites, and aqueous ozone).

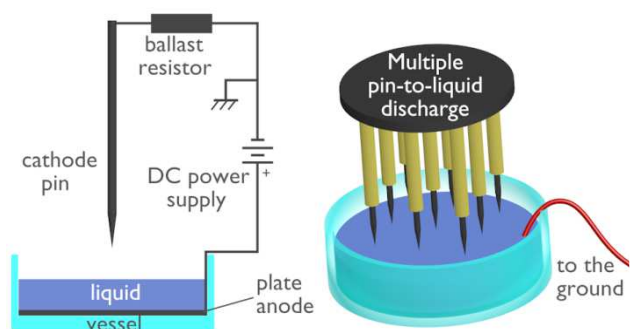

Figure 1. Schematics of the single pin-to-liquid discharge and of the multiple pins-to-liquid discharge.

## 3. Results & discussion

The **Figure 2** gives the V-I curve of a single pin-to-liquid discharge operating in ambient air. Four different stages of the discharge can be distinguished correlating electrical and optical measurements. In stage 1 (currents lower than 7 mA), the discharge behaves as in normal glow regime: the applied voltage is constant when increasing the current. A single but large discharge filament is formed, giving rise to a ring shape self-organized pattern on the liquid interface. In stage 2 (7-15 mA), the increase in the current at constant voltage makes the tungsten pin hotter, giving rise to a strong heat radiation as well as to very emissive nitrogen bands. The stage 3 (15-20 mA) corresponds to an unstable regime where the discharge switches between two states: a "high voltage" state (800-1000V) where it presents optical properties similar as those obtained in stage 2 and a "low voltage" state (600-700V) where the discharge optical properties are similar to those of stage 4. In that latter stage (>20 mA), the discharge recovers its stability while complex self-organized patterns are formed on the liquid interface. The typical patterns formed on this interface are shown in **Figure 3** for different current values.

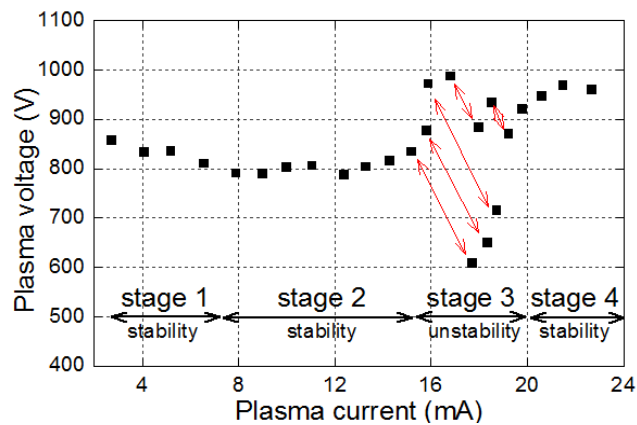

Figure 2. V-I characteristics of the pin-to-liquid electrode discharge.

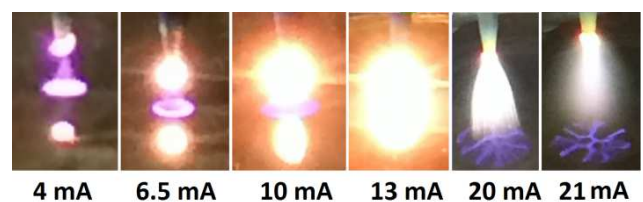

Figure 3. Pictures of the pin-to-liquid discharge above deionized liquid interface.

The **Figure 4** shows a detailed picture of the discharge interacting with liquid interface. A hot spot is clearly identified on the tungsten pin's tip while plasma filaments bridge the cathode pin to the liquid. On the liquid interface lie the plasma self-organized patterns. Filaments and patterns behavior have been characterized by switching DC power supply polarity, changing type of liquid and pin-to-liquid distance. Then the power deposited as well as number of filaments on a given residence time have been estimated. The temperature of these filaments a well as an electric field profile have been determined.

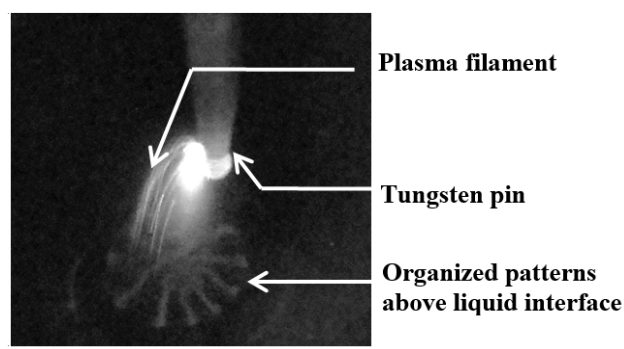

Figure 4. Self-organization patterns filaments resulting from a single pin-to-liquid discharge (stage 4)

In the treated liquids, reactive species such as hydrogen peroxide and nitrates have been quantified. A correlation between their concentration as well as plasma filaments and self-organized patterns characteristics is suggested.

Then, a multiple pins-to-liquid discharge has been designed to match with the expectations of plasma agronomy applications. Electrical and optical characterizations have been performed to compare electric field, gas temperature and the production of reactive species with those from single pin-to-liquid discharge. Also, how the filaments are now distributed and how they are connected with the self-organized patterns is discussed.

Finally, the plasma activated media prepared with the multiple pins-to-liquid discharge are used to irrigate lentils seeds. Their germination rate as well as the seedlings stems growth are daily recorded. They are compared to controls and to liquids activated by plasma jets to show the plus-value of the pins-to-liquid discharges in plasma agronomy applications.

## 4. Acknowledgements

This work has been done within the LABEX Plas@par project, and received financial state aid managed by the Agence Nationale de la Recherche, as part of the programme "Investissements d'avenir" under the reference ANR-11-IDEX-0004-02. This work has also been supported by the Île-de-France Region in the framework of the $^{PF2}$Abiomede Sesame project (16016309)